\documentclass[11pt,a4paper]{article}

\usepackage{amsfonts,dsfont}
\usepackage[T1]{fontenc}
\usepackage{cite}
\usepackage{bbm}
\usepackage{indentfirst}
\usepackage{amsopn}
\usepackage{amsmath}
\usepackage{amssymb}
\usepackage{amsthm}
\usepackage{authblk}
\usepackage{hyperref}
\usepackage{graphicx}
\usepackage{color}
\usepackage{listings}
\usepackage{mathtools}
\usepackage{subfig}
\usepackage{enumerate}
\usepackage{footnote}
\usepackage{physics}

\DeclareMathOperator{\Log}{Log}

\newcommand{\C}{\ensuremath{\mathbb{C}}}

\newcommand{\1}{{\rm 1\hspace{-0.9mm}l}}

\newtheorem{theorem}{Theorem}

\usepackage{caption}
\def\>{\rangle}
\def\<{\langle}

\newcommand{\appendixnumberline}[1]{Appendix\space}

\let\oldappendix\appendix
\makeatletter
\renewcommand{\appendix}{%
	\addtocontents{toc}{\let\protect\numberline\protect\appendixnumberline}%
	\renewcommand{\@seccntformat}[1]{Appendix~\csname the##1\endcsname:\quad}%
	\oldappendix
}
\makeatother

\begin{document}
	\title{Perturbation of the numerical range of unitary matrices}
	
	\author[1]{Ryszard Kukulski\footnote{rkukulski@iitis.pl}}
	\author[1]{Paulina Lewandowska}
	\author[1]{\L ukasz Pawela}
	\affil[1]{Institute of Theoretical and Applied Informatics, Polish 
		Academy of Sciences, ul. Ba{\l}tycka 5, 44-100 Gliwice, Poland}
	\date{}
	
	\maketitle
	
	\begin{abstract}
		In this work we show how to approach the problem of manimulating the 
		numerical range of a unitary matrix. This task has far-reaching impact 
		on the study of discrimination of quantum measurements. We achieve the 
		aforementioned manipulation by introducing a method which allows us to 
		find a unitary matrix whose numerical range contains the origin where 
		at the same time the distance between 
		unitary matrix and its perturbation is 
		relative small in given metric.
	\end{abstract}

	\section{Introduction}
	One of the most important tasks in quantum information theory is a problem 
	of 
	distinguishability of quantum 
	channels~\cite{helstrom1976quantum,jenvcova2014base}.  Imagine we have an 
	unknown
	device, a black-box. The only information we have is that it performs one 
	of 
	two channels, say $\Phi$ and $\Psi$. 
	We want 
	to tell whether it is possible to discriminate $\Phi$ and $\Psi$ perfectly, 
	i.e. with probability equal to one. Helstrom's 
	result~\cite{helstrom1969quantum} gives the analytical formula for upper 
	bound  of probability of discrimination quantum channels using the special 
	operators norm called diamond norm or sometimes referred to as completely 
	bounded trace norm~\cite{aharonov1998quantum}. The Holevo-Helstrom theorem 
	says that the quantum channels $\Psi$ and $\Phi$ are perfectly 
	distinguishable if and only if the distance between them is equal two by 
	using diamond norm.   In general, numerical computing of diamond norm is a 
	complex task. Therefore, researchers were limited to smaller classes of 
	quantum channels.  One  of  the  first  results  was  the  study  of  
	discrimination  of  unitary 
	channels $\Phi_U: \rho \rightarrow U \rho U^\dagger$ where $\rho$ is a 
	quantum 
	state.  The sufficient condition for perfect discrimination  of unitary 
	channels $\Phi_U $ and $\Phi_\1$ is that zero belongs to the numerical 
	range 
	of unitary matrix $U$~\cite{watrous2018theory}.

	The situation in which zero belongs to numerical range of unitary matrix 
	$U$ 
	paves the way toward simple calculating of probability of discrimination 
	unitary channels  without the necessity of  computing the diamond norm. Now 
	consider the following scenario.  
	We have two quantum channels $\Phi_U$ and $\Phi_\1$ such that zero does not 
	belong to the numerical range of $U$. Hence, we know that we cannot 
	distinguish between $\Phi_U$ and $\Phi_\1$ perfectly. Therefore, we can 
	assume some kind of noise and consider the unitary channel $\Phi_V$ beside 
	$\Phi_U$ such that the distance between unitary matrix $V$ and $U$ is 
	relative small where at the same time zero belongs in numerical range of 
	$V$.  Such a unitary matrix $V$ will be called perturbation of $U$. 
	
	In this work we are interested in  determining the perturbation form of 
	$V$. Our motivation is two-fold. On the one hand considering the unitary 
	channels $\Phi_V$ and $\Phi_\1$ we know that they  will be perfectly 
	distinguishable. On the other hand our method of computing $V$ does not 
	change the measurement result in standard basis. 
	
	Our work is naturally divided into three parts. In the first part we show 
	the
	mathematical preliminaries needed to present our main result.  The second 
	part presents
	the theorem which gives us the method of manipulation of numerical range of 
	unitary matrices. In third part we show the example illustrative our 
	theorem.   Concluding  remarks  are presented in the end of our work. 
	
	\section{Mathematical preliminaries}
	Let us introduce the following notation. Let $\C^d$ be complex  $d$-dimensional vector space. We denote the set of all matrix 
	operators  by 	$\mathrm{L}(\C^{d_1}, \C^{d_2})$ while the set of isometries by 
	$\mathrm{U}(\C^{d_1}, \C^{d_2}) $. 
	It easy to see that every square isometry is a unitary matrix. The set of all 
	unitary matrices we will be denoted by $ U \in \mathrm{U}(\C^d)$.   We will be 
	also interested in diagonal matrices and diagonal unitary matrices denoted by 
	$\mathrm{Diag}(\C^d)$ and $\mathrm{DU}(\C^d)$ respectively. Next classes of 
	matrices that will be used in this work are Hermitian matrices denoted by 
	$\mathrm{Herm}(\C^d)$.   All of the above-mentioned matrices are normal 
	matrices  i.e. $AA^\dagger = A^\dagger A$. Every normal matrix $A$ can be 
	expressed as a linear combination of projections onto pairwise orthogonal 
	subspaces
	\begin{equation}
	A = \sum_{i=1}^{k} \lambda_i \ketbra{x_i}{x_i},
	\end{equation}
	where scalar $\lambda_i \in \C$ is an eigenvalue of $A$ and $\ket{ x_i} \subset 
	\C^d$ is an eigenvector corresponding to the eigenvalue $\lambda_i$. This 
	expression of a normal matrix $A$ is called a spectral decomposition of $A$ 
	~\cite{hall2013quantum}.  Many interesting and useful norms, not only for 
	normal matrices,  can be defined on spaces of matrix operators. In this work we 
	will mostly be concerned with a family of norms called Schatten~\cite{franklin2012matrix}
	p-norms defined as
	\begin{equation}
	||A||_p = \left( \tr \left( \left( A^\dagger A \right)^{\frac{p}{2}} \right) \right)^\frac{1}{p}
	\end{equation}
	for any $A \in \mathrm{L}(\C^{d_1}, \C^{d_2})$.
	The Schatten $\infty$-norm is defined as 
	\begin{equation}
	||A||_\infty = \max \left \{  ||A \ket{u} || : \ket{u} \in \C^{d_1} , ||\ket{u} 
	|| \le 1 \right\}.
	\end{equation}
	For a given square matrix $A$  the set of all 
	eigenvalues of $A$ will be denoted  by $\lambda(A)$ and  $r(\lambda_i) $ will denote the multiplicity of each eigenvalue $\lambda_i \in \lambda(A)$.
	For any square matrix $A$, one defines its numerical range~\cite{murnaghan1932field,gustafson1997field} as a subset of the complex plane 
	\begin{equation}
	W(A) = \{ z \in \C: z = \bra{\psi} A \ket{\psi}, \ket{\psi} \in\C^d, \braket{\psi}{\psi} = 1 \}.
	\end{equation}
	It is easy to see that $\lambda(A) \subseteq W(A)$. One of 
	the most important properties of $W(A)$ is its convexity which was shown by 
	Hausdorff and Toeplitz~\cite{toeplitz1918algebraische,hausdorff1919wertvorrat}. For any normal matrix $A$ the set $W(A)$ is a convex 
	hull of spectrum of $A$ which will be denoted by $\mathrm{conv}(\lambda(A))$. 
	Another well-known property of $W(U)$ for any unitary matrix $U \in 
	\mathrm{U}(\C^d)$ is the fact that  its numerical range forms a polygon whose 
	vertices are eigenvalues of $U$ lying in unit circle on complex plane. In 
	our work we introduce the counterclockwise order of eigenvalues of unitary 
	matrix $U$ ~\cite{bhatia2013matrix} such that we choose any eigenvalue 
	named $\lambda_1 \in \lambda(U)$ on the unit circle and next eigenvalues 
	are labeled counterclockwise.
	
	In our setup we consider the space $\mathrm{L}(\C^d)$. Imagine that the 
	matrices are points in space $\mathrm{L}(\C^d)$ and the 
	distance between them is bounded by small constant $0 < c \ll 1$. 
	We will take two unitary matrices - matrix $U \in \mathrm{U}(\C^d)$ and its 
	perturbation $V \in \mathrm{U}(\C^d)$ i.e. $||U-V||_\infty \le c$ by 
	using 
	$\infty$-Schatten norm. We want to determine the path connecting these points 
	given by smooth curve.  To do so, we fix continuous parametric (by 
	parameter $t$) 
	curve $U(t) \in \mathrm{U}(\C^d)$ for any $t \in [0,1]$ with boundary 
	conditions $U(0) := U$ and $U(1) := V $. The most natural and also the shortest 
	curve connecting $U$ and $V$ is geodesic~\cite{antezana2014optimal} given 
	by
	\begin{equation}
	t \rightarrow U \exp\left(t \Log \left(U^\dagger V\right)\right), 
	\end{equation}
	where $\Log$ is the matrix function such that it changes eigenvalues 
	$\lambda \in 
	\lambda(U)$ into $\log(\lambda(U))$, where $-i\log(\lambda(U)) \subset 
	(-\pi, 
	\pi]$.
	
	We will study how the numerical range $W(U(t))$ will be changed depending on parameter 
	$t$. Let $H := -i \Log \left( U^\dagger V \right)$. Let us see that $H \in 
	\mathrm{Herm}(\C^d)$ and $W(H) \subset (-\pi, \pi]$ for any $U,V \in 
	\mathrm{U}(\C^d)$.  We can also observe that 
	\begin{equation}
	\begin{split}
	W\left(U\exp\left(itH\right)\right) &= W\left( U \exp \left( it VDV^\dagger 
	\right) \right) = W\left( UV \exp \left( it D \right) V^\dagger 
	\right) \\& = W\left( V^\dagger U V \exp \left( it D \right) \right) = W\left( 
	\widetilde{U} \exp \left( it D\right) \right)
	\end{split}
	\end{equation} where $\widetilde{U} := V^\dagger U V \in \mathrm{U}(\C^d)$. 
	Hence, without loss of generality we can assume that $H$ is a diagonal matrix. 
	Moreover, we can assume that $D \ge 0$ which follows from simple calculations
	\begin{equation}
	\begin{split}
	W \left( U \exp \left(it D \right) \right) &= W \left( U \exp \left(it D_{+} \right) \left( \exp \left( it \alpha \1 \right) \right) \right) = W \left( e^{it \alpha} U \exp \left(it D_{+} \right) \right) \\&= W \left( U \exp \left(it D_{+} \right) \right).
	\end{split}
	\end{equation} 
	Let us see that the numerical range of $U(t)$ for any $t \in [0,1]$ is 
	invariant to above calculations although the trajectory of $U(t)$  is changed. 
	Therefore, we will consider the curve
	\begin{equation}
	t \rightarrow U \exp \left( it D_{+} \right),
	\end{equation}
	where $t \in [0,1]$ and $U \in \mathrm{U}(\C^d)$, $D_+ \in \mathrm{Diag}(\C^d)$ 
	such 
	that $D_+ \ge 0$.

	%
	\section{Main result}
	In this section we will focus on the 
	behavior of the spectrum of the unitary matrices $U(t)$, which will reveal the 
	behavior of $W(U(t))$ for relatively small parameter $t$.  Without loss of 
	generality we can assume that $\tr \left( D_+ \right)=1$. Together with the fact that $D_+ \in \mathrm{Diag}(\C^d)$ and $D_+ \ge 0$ we can note that  $D_+ = \sum_{i=1}^{d} p_i \ketbra{i}{i}$, where $p \in \C^d$ is a probability vector.	Let us also define the set
	\begin{equation}
	S_\lambda^M=\left\{\ket{x} \in \C^d: 
	(\lambda\1_d-M)\ket{x}=0, \|\ket{x}\|_2=1\right\}
	\end{equation}
	for some matrix $M \in \mathrm{L}(\C^d)$ which consists of unit eigenvectors corresponding to the eigenvalue $\lambda$ of 
	the 
	matrix $M$. We denote by $k=r(\lambda)$ the multiplicity of eigenvalue 
	$\lambda$ whereas by 
	$I_{M,\lambda} \in 
	U(\C^k,\C^d)$ we denote the isometry which columns are formed by eigenvectors corresponding to eigenvalue 
	$\lambda$ of a such
	matrix $M$. Let $\lambda(t), \beta(t) \in \C$ for $t \ge 0$. We will write 
	$\lambda(t) \approx \beta(t)$ for relatively small $t \geq 0$, whenever 
	$\lambda(0)=\beta(0)$ and 
	$\frac{\partial}{\partial 
		t}\lambda(0)=\frac{\partial}{\partial 
		t}\beta(0)$.
	\begin{theorem}\label{tw}
		Let $U \in \mathrm{U}(\C^d)$ be a
		unitary matrix with spectral decomposition 
		\begin{equation}
		U=\sum_{j=1}^d \lambda_j \ketbra{x_j}.
		\end{equation}	 Assume that the eigenvalue
		$\lambda \in \lambda(U)$ is such that
		$r(\lambda) = k$. 		Let us define a matrix $V(t)$ given by 
		\begin{equation}
		V(t) = \exp(itD_+)=\sum_{i=1}^{d} e^{i p_i t} \ketbra{i} \in 
		\mathrm{DU}(\C^d), \quad t \geq 0.
		\end{equation}
		Let $\lambda(t):=\lambda(UV(t)) $  and let every $\lambda_j(t) \in 
		\lambda(t)$ corresponds to eigenvector 
		$
		\ket{x_j(t)} $. Assume that $\lambda_{1}(t), 
		\ldots, \lambda_{k}(t)$ are such eigenvalues that $\lambda_{j}(t) 
		\to 
		\lambda$, as $t \to 0$. Then:
		\begin{enumerate}[(a)]
			\item If $\min\limits_{\ket{x} \in S_\lambda^U} \sum\limits_{i=1}^d\ 
			p_i 
			|\braket{i}{x}|^2=0$, then $\lambda$ is an eigenvalue of $UV(t)$.
			\item If $|\{p_i: p_i>0\}|=l<k$, then 
			$\lambda$ is an eigenvalue of $UV(t)$ and $r(\lambda) \ge
			k-l.$
			\item Each eigenvalue of product $UV(t)$ moves counterclockwise or 
			stays in the initial position as 
			parameter $t$ increases.
			\item If $k=1$, then
			\begin{equation*}
			\lambda_{1}(t) \approx \lambda \exp\left( i t 
			\sum\limits_{i=1}^d\ 
			p_i |\braket{i}{x_1}|^2 \right)
			\end{equation*}
			for small $t \geq 0$.
			\item Let $Q:=I_{U,\lambda}^\dagger D_+ I_{U,\lambda}$ and
			$\lambda_1(Q) \le \lambda_2(Q) \le \ldots \le \lambda_k(Q) $. Then we have
			\begin{equation*}
			\lambda_{j}(t) \approx \lambda \exp\left( i \lambda_{j}(Q) t 
			\right)
			\end{equation*}
			for small $t \geq 0$ and eigenvector $\ket{ x_j}$ corresponding 
			to $\lambda_j \in \lambda(U)$  is given by 
			\begin{equation*}
	 \ket{x_j}=I_{U,\lambda} \ket{ v_j},			\end{equation*} where 
	 $\ket{ v_j} \in 
			S^{Q}_{\lambda_j(Q)}$.		
			\item For each $j=1,\ldots,d$ we have
			\begin{equation*}
			\frac{\partial}{\partial 
				t}\lambda_j(t)=i \lambda_j(t)\sum_{i=1}^d p_i 
			|\braket{i}{x_j(t)}|^2.
			\end{equation*}
			Moreover, \begin{equation*}
			\sum_{j=1}^d \left|\frac{\partial}{\partial 
				t}\lambda_j(t)\right|=1.
			\end{equation*}
		\end{enumerate}
	\end{theorem}

This theorem gives us equations which one can use to predict behavior of 
$W(UV(t))$. Observe the postulate $(f)$ fully determines the movement of the 
spectrum. However, this is a theoretical statement and in practice determining 
the function $t \mapsto \ket{x_j(t)}$ is a numerically complex task. The 
postulates $(a)-(e)$ play a key role in numerical calculations of $W(UV(t))$. 
The most important fact comes from $(c)$ which says that all eigenvalues move 
in 
the same direction or stay in the initial position. The instantaneous velocity 
of a given eigenvalue in general case is given in $(e)$, while in the case of 
eigenvalue with multiplicity equal one, the instantaneous velocity is 
determined by $(d)$. We see that whenever the spectrum of the matrix $U$ is not 
degenerated, calculating these velocities is easy. What is more, when some 
eigenvalue is 
degenerated, the postulate $(e)$ not only gives us method to calculate the 
trajectory of this eigenvalue, but also determines the form of corresponding 
eigenvector. It is worth noting that the postulates $(d),(e)$ give us 
only an
approximation of the velocities, so despite being useful in numerical calculations, 
these expressions are valid only in the neighborhood of $t=0$. Moreover, 
sometimes we are able to precisely specify this velocities. This happens in the 
cases presented in $(a),(b)$. Whenever the calculated velocity is zero we know 
for 
sure that this eigenvalue will stay in the initial position. According to the 
postulate $(b)$ the same happens when the multiplicity of the eigenvalue is 
greater than the number of positive elements of vector $p$.

	\section{Example}
\begin{figure}[h]
			\centering
				\includegraphics[width=\textwidth]{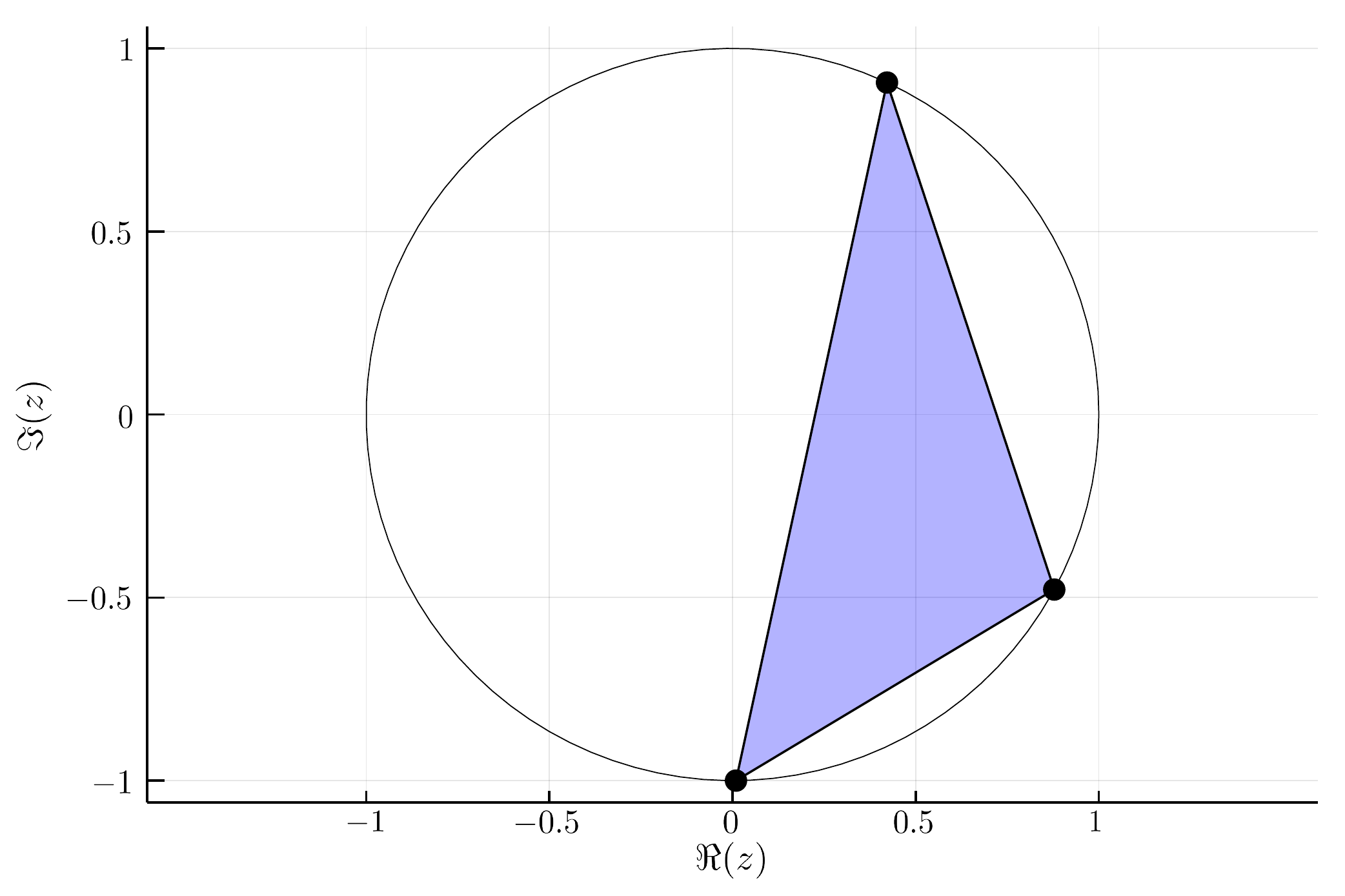}
				\caption{The numerical range $W(U)$.}
				\label{f-1}
		\end{figure}

	We start with sampling some random unitary matrix $U \in \mathrm{U}(\C^3)$ 
	such that $0 \not \in W(U)$
	\begin{equation}
	U= 
	\left[\begin{matrix}
	0.267868+0.026891i &    0.752935-0.510663i & -0.314404-0.0313982i \\
	-0.83413-0.0693252i &  0.245915-0.275811i &   0.34174-0.214685i \\
	0.472125+0.0635826i & 0.0211772-0.18793i  &  0.795835-0.322391i   
	\end{matrix}\right]
	\end{equation}
	for which numerical range is given in Figure \ref{f-1}.
Next, we can calculate eigenvectors of matrix $U$ and according to the 
postulate $(d)$ choose appropriate probability vector $p$. The squared modules 
of eigenvectors entries form the matrix  
	\begin{equation}
	Q=\left[\begin{matrix}
	0.426542& 0.543517& 0.0299407\\
	0.0480551& 0.105588& 0.846357\\
	0.525403& 0.350895& 0.123702
	\end{matrix}\right]
	\end{equation}
	Rows of the matrix $Q$ correspond to the considered eigenvectors. Here, we 
	will 
	focus on the most distant pair of eigenvalues $(\lambda_1,\lambda_3)$ and 
	their eigenvectors which are given in the first and the third row, 
	respectively. As we 
	can see the greatest difference in speed is in the second column, namely 
	between values $Q_{1,2}= 0.543517$ and $Q_{3,2}= 0.350895$. That means we 
	would like to rotate this spectrum clockwise. To do so, we consider vector 
	$p=(0,1,0)$ and change the direction from counterclockwise to clockwise by
	taking matrix $V(t)^\dagger$ instead of $V(t)$.
	
\begin{figure}[h]
		\centering
			\includegraphics[width=\textwidth]{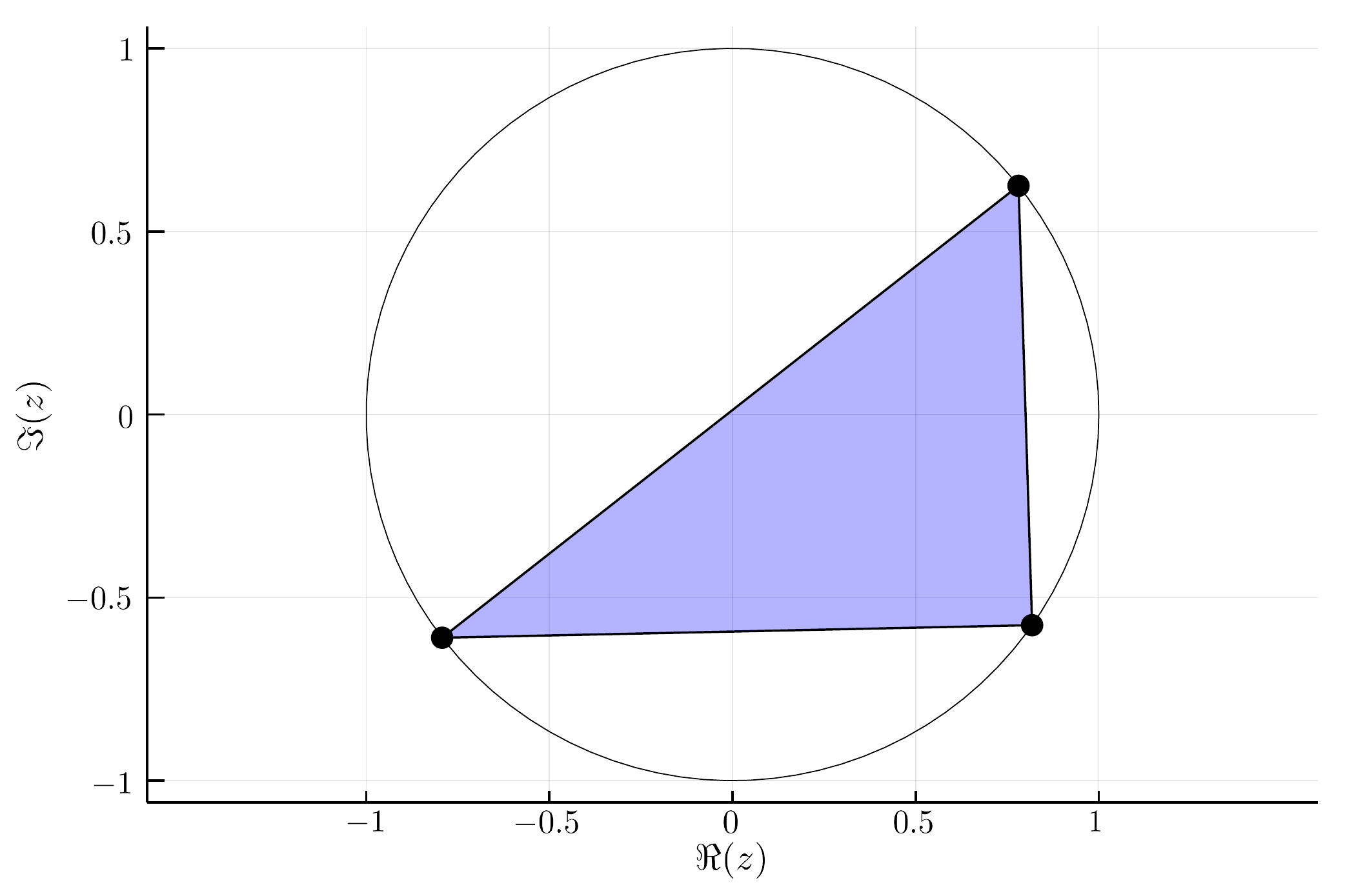}
			\caption{The numerical range $W(UV(1.5)^\dagger)$.}
			\label{f-2}
	\end{figure}

	The numerical range of matrix $UV(t)^\dagger$ after time $t=1.5$ is given 
	in Figure~\ref{f-2}. 
We can see that $0 \in W(UV(1.5)^\dagger)$, but numerical calculations show 
that $0 \in W(UV(t)^\dagger)$ for $t \approx 1.45$.
	\section{Conclusion and discussion}
	In this work we considered an approach to manipulation of the numerical 
	range of unitary matrices. That was done by multiplying given unitary 
	matrix $U$ by some unitary matrix $V$ which is diagonal in the fixed 
	computational basis (we took the standard basis) and is relatively close to 
	identity matrix. We established differential equations describing behavior 
	of eigenvalues and presented their approximated solutions, which we find 
	useful in numerical calculations. Our motivation was to find for given 
	unitary matrix $U$ the closest unitary matrix of the form $UV$ such that 
	channels $\Phi_{UV}$ and $\Phi_\1$ are perfectly distinguishable. It is 
	important to stress that applying channel $\Phi_V$ to the quantum states 
	leaves their classical distribution unchanged.  
	
	The results in Theorem $\ref{tw}$ are suitable to solve 
	various tasks. For example, one would like to maximize 
	the distance between the numerical range and the point zero. Such task was 
	introduced in \cite{puchala2018strategies} and plays crucial role in 
	calculating diamond norm of difference of two von Neumann measurements.
	\section*{Acknowledgements}
	
	This work was supported by the Foundation for Polish Science (FNP) under grant 
	number POIR.04.04.00-00-17C1/18-00.
	
	\bibliographystyle{ieeetr}
	\bibliography{unitary}
	
	\appendix
	\section{Proof of Theorem \ref{tw}}
		\begin{proof}
		$(a)$ This fact is implicated by equation $\sum\limits_{i=1}^d\ 
		p_i 
		|\braket{i}{x}|^2=0$ for some eigenvector $\ket{x}$ of eigenvalue 
		$\lambda$. Eventually, we obtain $UV(t)\ket{x}=U \ket{x} = \lambda \ket{x}$.
		
		$(b)$ We will show that there are at least $k-l$ orthogonal 
		eigenvectors 
		$\ket{x}$ of 
		eigenvalue 
		$\lambda$, for which $\sum\limits_{i=1}^d\ 
		p_i 
		|\braket{i}{x}|^2=0$, so $(a)$ will imply $(b)$. W.l.o.g. 
		assume $p_i>0$ for $i=1,\ldots,l$. For each matrix $W \in U(\C^k)$ the 
		columns 
		of isometry $I_{U,\lambda}W$ consist of eigenvectors of eigenvalue 
		$\lambda$. 
		We can choose 
		such a matrix $W$ for which first $k-l$ columns are orthogonal to each 
		of 
		vectors $I_{U,\lambda}^\top \ket{1}, \ldots, I_{U,\lambda}^\top 
		\ket{l}$. One can 
		note that for $\ket{x} 
		\in \{ I_{U,\lambda}W\ket{1},\ldots,I_{U,\lambda}W \ket{k-l} \}$ we obtain 
		$\sum\limits_{i=1}^d\ 
		p_i 
		|\braket{i}{x}|^2=0$.
		
		$(c)$ Fix some eigenvalue $\lambda$ with $r(\lambda)=k$. We introduce 
		the notation of 
		$\Pi=I_{U,\lambda}I_{U,\lambda}^\dagger$. We consider the subspace 
		$\{\ket{x}: D_+\ket{x}=0, \Pi\ket{x}=\ket{x}\}$ along with projection 
		$\Pi_0$ on this subspace. Denote $\Pi_+=\Pi-\Pi_0$. Let
		\begin{equation}
		c:=\min_{\ket{x}: \Pi_+\ket{x}=\ket{x},\|\ket{x}\|_2=1} 
		\bra{x}D_+\ket{x}.
		\end{equation}
		One can note that $c>0$.
		Take unit vector $\ket{x}$ and define $\ket{v}=(\1_d-\Pi)\ket{x}$. 
		First of all, we will show that 
		$\tr (\1_d-\Pi) \ketbra{x} \in O(\epsilon)$ if 
		$|\lambda-\bra{x}U\ket{x}|=\epsilon$. Direct calculations reveal that
		\begin{equation}
		\begin{split}
		\epsilon&=|\lambda-\bra{x}U\ket{x}|=|\lambda \tr \ketbra{x}-\tr U \Pi 
		\ketbra{x}-\tr U (\1_d-\Pi)	\ketbra{x}|\\
		&=|\lambda \tr (\1_d-\Pi) \ketbra{x}-\tr(\1_d-\Pi)U(\1_d-\Pi) \ketbra{x}|.
		\end{split}
		\end{equation}
		If $\|v\|=0$ the statement is true, 
		so assume $\|v\|\not=0$. We obtain
		\begin{equation}
		\epsilon=|\lambda \braket{v}-\bra{v}U\ket{v}|\geq \braket{v} 
		\mathrm{dist}(\lambda,\mathrm{conv}(\lambda_{k+1},\ldots,\lambda_d))>0,
		\end{equation}
		which finishes this part of proof.
		
		In the second part we will check the behavior of points $\bra{x}U\ket{x}$ 
		in the neighborhood of point $\lambda$. We assume that 
		$|\lambda-\bra{x} U \ket{x}| = \epsilon$ for relatively small 
		$\epsilon\geq 0$, so $\tr (\1_d-\Pi) \ketbra{x} 
		\in O(\epsilon)$. The derivative of 
		trajectory of 
		such a point is
		\begin{equation}
		\frac{\partial}{\partial t} 
		(\bra{x}UV(t)\ket{x})(0)=i \bra{x}UD_+\ket{x}.
		\end{equation}
		We can rewrite the above as
		\begin{equation}
		\begin{split}
		i \bra{x}UD_+ \ket{x} &= i \bra{x}(\Pi_++\Pi_0+\1_d-\Pi)UD_+ 
		(\Pi_++\Pi_0+\1_d-\Pi)\ket{x}=\\
		&=i \lambda \bra{x} \Pi_+ D_+ \Pi_+ \ket{x}+i\bra{x} \Pi_+ U D_+ 
		(\1_d-\Pi) \ket{x}\\
		&+i\bra{x} (\1_d-\Pi) U D_+ \Pi_+ \ket{x}+i\bra{x} (\1_d-\Pi)U 
		D_+ (\1_d-\Pi)\ket{x}).
		\end{split}
		\end{equation}
		The above equation means that the instantaneous velocity of point 
		$\bra{x}U\ket{x}$ is the sum of the velocity
		\begin{equation}
		i \lambda \bra{x} \Pi_+ D_+ \Pi_+ \ket{x}
		\end{equation}
		which is responsible for counterclockwise movement and which speed is $ 
		\bra{x} \Pi_+ D_+ \Pi_+ \ket{x}$ and the ``noise'' velocity which for 
		the 
		most pessimistic scenario can be rotated in any direction and which 
		speed is at most
		\begin{equation}
		|\bra{x} \Pi_+ U D_+ 
		(\1_d-\Pi) \ket{x}|+|\bra{x} (\1_d-\Pi) U D_+ \Pi_+ \ket{x}|+|\bra{x} 
		(\1_d-\Pi)U 
		D_+ (\1_d-\Pi)\ket{x}|.
		\end{equation} 
		The speed of velocity with direction $i \lambda$ can be lower bounded by
		\begin{equation}
		c \bra{x} \Pi_+ \ket{x},
		\end{equation}
		while the speed of the second velocity can be upper bounded by
		\begin{equation}
		2\sqrt{\bra{x} \Pi_+ \ket{x}}\sqrt{\bra{x} (\1_d-\Pi) \ket{x}}+\bra{x} 
		(\1_d-\Pi) \ket{x} \leq 2\sqrt{\bra{x} \Pi_+ 
		\ket{x}}\sqrt{\epsilon}+\epsilon.
		\end{equation}
		There exists constant $d > 0$ depending on the geometry of the 
		numerical range of the matrix $U$, such that if
		\begin{equation}
		c \bra{x} \Pi_+ \ket{x} \geq d(2\sqrt{\bra{x} \Pi_+ 
			\ket{x}}\sqrt{\epsilon}+\epsilon),
		\end{equation}
		then the point $\bra{x} U \ket{x}$ moves counterclockwise. This is true 
		if
		\begin{equation}
		\bra{x} \Pi_+ \ket{x} \geq \frac{cd+2d^2+2d \sqrt{d^2+cd}}{c^2} 
		\epsilon.
		\end{equation}
		In the case when the above inequality does not hold, the speed 
		of the second velocity is upper bounded by a some linear function of 
		variable $\epsilon$. That means there exists $t_0$ such that for $t 
		\leq t_0$ there can not exists eigenvalue $\widetilde{\lambda} \in 
		UV(t)$ which $\widetilde{\lambda} \to \lambda$ as $t \to 0$ and 
		$\widetilde{\lambda}$ is before $\lambda$ in the counterclockwise order.
		To finish the proof we can see the above holds for any $t \geq 0$ due 
		to fact that $V(t_0 + t)=V(t_0)V(t)$.

		$(d)$ To see this we first need to describe local dynamics of 
		point $\beta(t)=\bra{x_1}UV(t)\ket{x_1}= \lambda \bra{x_1}V(t) 
		\ket{x_1}.$ One can note that $\beta(0)= \lambda$ and 
		$\frac{\partial}{\partial t} \beta (0)=i \lambda \sum\limits_{i=1}^d	
		p_i |\braket{i}{x_1}|^2 $. That means $\beta(t) \approx \lambda \exp(it 
		\sum_{i=1}^d p_i |\braket{i}{x_1}|^2)$. To see that $\beta(t) \approx 
		\lambda_1(t)$ we need to utilize the following facts:
		\begin{itemize}
			\item Eigenvalues of $UV(t)$ are continuous functions.
			\item $\beta(t) \in W(UV(t)).$
			\item Trajectory of $\beta(t)$ is curved in such a way that holds 
			$\frac{1-|\beta(t)|}{|\beta(t) - \lambda|} \rightarrow 0.$
		\end{itemize}
			The above means that if $R(t) \subset \{|z|=1\}$ is an 
			arch in which we can potentially find eigenvalue $\lambda_1(t)$ 
			according to the fact that 
		$\beta(t) \in 
		W(UV(t))$, then 
		it is true that $\frac{|R(t)|}{|\beta(t)-\lambda|} 
		\to 0$ and consequently $\lambda_1(t) \approx \beta(t)$ for small $t 
		\geq0$.

		$(e)$ The see this point we need to utilize the postulate $(c)$ along 
		with the proof of the postulate 
		$(d)$ for eigenvectors $\ket{x} \in \{ I_{U,\lambda} \ket{ v_1}, 
		\ldots, I_{U,\lambda} \ket{v_k} \}$, where $Q:=I_{U,\lambda}^\dagger 
		D_+ I_{U,\lambda}$ and $\ket{ v_j} \in S^{Q}_{\lambda_j(Q)}$ for 
		$j=1,\ldots,k$.
		
		$(f)$ This relation follows from $(e)$.
	\end{proof}
\end{document}